\documentstyle[prl,aps,floats]{revtex}
\begin{document}
\draft
%%%%%%%%%%%%%%%%%%%%%%%%%%%%%%%%%%%%%%%%%%%%%%%%%%%%%%%%%%%%%%%%%%%%%%
%
%  Uncomment following four lines and one below for 2 column format
%  and figure insertions.
%
\input epsf
\renewcommand{\topfraction}{1}
\twocolumn[\hsize\textwidth\columnwidth\hsize\csname
@twocolumnfalse\endcsname
%%%%%%%%%%%%%%%%%%%%%%%%%%%%%%%%%%%%%%%%%%%%%%%%%%%%%%%%%%%%%%%%%%%%%%
\preprint{hep-ph/9802306}
\title{The formation rate of semilocal strings}
\author{Ana Ach\'{u}carro,$^1$ Julian Borrill~$^2$ and Andrew R.~Liddle~$^3$}
\address{~\\
$^1$Department of Theoretical Physics, UPV-EHU, Bilbao, Spain\\
$^1$Institute for Theoretical Physics, University of Groningen, The 
Netherlands\\
$^2$Center for Particle Astrophysics, University of California,
Berkeley, CA 94720\\ $^2$National Energy Research Scientific Computing
Center, Lawrence Berkeley National Laboratory, University of
California, Berkeley, CA 94720 \\
$^3$Astronomy Centre, University of Sussex, Falmer, Brighton BN1 
9QJ, Great Britain\\
$^3$Astrophysics Group, The Blackett Laboratory, Imperial College, London SW7 
2BZ, 
Great Britain (present address)}
\date{\today}
\maketitle
\begin{abstract}
We carry out three-dimensional numerical simulations to investigate
the formation rate of semilocal strings. We find that the
back-reaction of the gauge fields on the scalar field evolution is
substantial, and leads to a significant formation rate in the
parameter regime where the semilocal strings are classically
stable. The formation rate can be as large as a third of that for
cosmic strings, depending on the model parameters.
\end{abstract}
\pacs{PACS numbers: 11.27.+d, 11.15.Ex \hspace*{6.6cm} hep-ph/9802306}

%  This is the other line to be uncommented for 2 column format
\vskip2pc]
%%%%%%%%%%%%%%%%%%%%%%%%%%%%%%%%%%%%%%%%%%%%%%%%%%%%%%%%%%%%%%%%%%%%%%

\section{Introduction}

It is expected that the early Universe underwent a series of phase transitions
as it cooled down.  Typically, such transitions lead to the formation of a
network of defects \cite{K76}, which may be of relevance to a number of
phenomena including structure formation in the Universe and
baryogenesis~\cite{reviews}.  The best known examples are topological defects
such as cosmic strings, whose stability is guaranteed by the topological
structure of the symmetry breaking at the phase transition, but it is also
possible for non-topological defects to form.  An example of the latter is
semilocal strings \cite{VA,Mark}; the semilocal string model is simply the
Weinberg--Salam model without fermions or W~bosons, in the limit in which the
SU(2) coupling constant is set to zero.  The only parameter in the theory is
$\beta \equiv m_{{\rm s}}^2/m_{{\rm v}}^2$, the ratio between the scalar and
vector masses (squared).  Its vacuum manifold is the three-sphere $S^3$, and has
no non-contractible loops.  Despite this Nielsen--Olesen vortices \cite{NO} may
form, and are classically stable if $\beta <1$ \cite{VA,Mark}.  Semilocal
strings are closely related to electroweak strings \cite{EW}, which can be
formed in the electroweak phase transition, and so understanding the formation
and evolution of these non-topological defects is an important task.

The semilocal model is characterized by the gauge fields having insufficient
degrees of freedom to be able to completely cancel the scalar field gradients,
even away from the core of any strings which form.  While their stability
depends on the parameter $\beta$, {\it stable} semilocal strings are stable not
only to small perturbations but also to semiclassical tunnelling \cite{PV}.
Further, although they can, unlike topological strings, come to an end (in what
is effectively a global monopole), they will not decay by breaking into smaller
segments \cite{GORS}.  On the contrary, Hindmarsh \cite{H} has conjectured that
the long-range interaction between these monopoles should lead to short pieces
of strings growing into longer ones.

We have recently shown~\cite{ABL97}, using a toy model with parallel strings, 
that semilocal strings can be identified by
studying the pattern of magnetic flux in a simulation.  For topological strings,
an estimate of the formation rate in systems with planar symmetry is sufficient
to determine the three-dimensional rate, since topology prevents the strings
from having an end \cite{VV}.  But non-topological strings can terminate, with 
the flux spreading out.  The closest to a three-dimensional
analytic estimate for semilocal strings is Hindmarsh's calculation \cite{H} of
the average magnetic flux through a correlated area at $\beta = 0$, with the
conclusion that vortices are rare.  For the electroweak string, Nagasawa and
Yokoyama \cite{NY} proposed a technique based on studying the scalar field alone
and concluded that the initial density would be negligibly small.  However, both
approaches neglect the gauge fields, which we have found to play a key
role~\cite{ABL97}.

In this paper we aim to estimate the density of semilocal strings at formation.
The usual argument for the formation of (topological) cosmic strings relies on
the vacuum manifold having non-contractible loops which can force the existence
of closed or infinite lines in which the Higgs scalar must have zero value,
confining the magnetic field to these vortex structures where the symmetry has
been restored.  The lack of topology prevents such arguments being employed for
semilocal strings, and numerical methods must be employed.  This requires two
parts --- a plausible initial configuration where the field configuration
captures the essence of a thermal phase transition (principally, the existence
of a correlation length) and secondly dynamical evolution to allow the strings
to `condense out' and be counted.   The numerical simulation of a network of
defects is a difficult problem because of the large range of scales involved in
the dynamics, but it has come within the capability of modern supercomputers. It 
is often tacitly assumed that only topological defects are
sufficiently robust to form in a phase transition through the Kibble
mechanism \cite{K76}.
Given that it is not presently known whether semilocal strings form at a
comparable density to cosmic strings, or with a completely negligible density,
our target is an order-of-magnitude estimate.

\section{The simulations}

We work in flat space-time throughout. The Lagrangian for the simplest
semilocal string model \cite {VA} is
\begin{eqnarray}
\label{eL}
{\cal L} & = & \left( \partial_\mu - i A_\mu \right) \phi_1^\dagger \left(
        \partial^\mu + i A^\mu \right) \phi_1 \nonumber \\
 & & + \left( \partial_\mu -
        i A_\mu \right) \phi_2^\dagger \left( \partial^\mu + i A^\mu \right)
        \phi_2 \nonumber \\
 & & - \frac{1}{4} F_{\mu\nu} F^{\mu\nu} - 
\frac{\beta}{2} \left( |\phi_1|^2 + |\phi_2|^2 - 1 
        \right)^2 \,,
\end{eqnarray}
where $\phi_1$ and $\phi_2$ are two equally-charged complex scalar
fields, $A_\mu$ is a U(1) gauge field and $F_{\mu\nu}$ the associated
gauge field strength.  Notice that the gauge coupling and the vacuum
expectation value of the Higgs have been set to one by choosing
appropriate units (the inverse vector mass, for length, and the
symmetry breaking scale, for energy).  The only remaining parameter in
the theory is $\beta = m_{{\rm s}}^2 / m_{{\rm v}}^2$, whose value
determines the stability of an infinitely long, straight, semilocal
string with a Nielsen--Olesen profile: it is stable
for $\beta < 1$, neutrally stable for $\beta = 1$ and unstable for
$\beta > 1$ \cite{VA,Mark,AKPV}. For $\beta = 1$ there is a family of
solutions with the same energy and different core widths, of which
only the semilocal string has complete symmetry restoration in the
center \cite{Mark}.

We work in temporal gauge $A_0=0$. Splitting the scalar fields into
four real scalars via $\phi_1 = \psi_1 + i \psi_2$, $\phi_2 = \psi_3 +
i \psi_4$, the equations of motion are
\begin{eqnarray}
\ddot{\psi}_a - \nabla^2 \psi_a + \beta \left( \psi_1^2 +
        \psi_2^2 + \psi_3^2 + \psi_4^2 - 1 \right) \psi_a \\
 \hspace*{1cm} + A^2 \psi_a + (-1)^{b} \left( 2 A.\nabla + \nabla.A \right) 
        \psi_b = 0  \,, \nonumber
\end{eqnarray}
(where $b$ is the complement of $a$ --- $1 \leftrightarrow 2$, $3
\leftrightarrow 4$ and dots are time derivatives) for the
scalar fields and
\begin{eqnarray}
\ddot{A}_i - \nabla^2 A_i + \partial_i \nabla.A + 2 \left(
        \psi_1 \stackrel{\leftrightarrow}{\partial_i} \psi_2 +
        \psi_3 \stackrel{\leftrightarrow}{\partial_i} \psi_4 \right) \\
 + 2 A_i \left( \psi_1^2 + \psi_2^2 + \psi_3^2 + \psi_4^2 \right)
        = 0 \,, \nonumber
\end{eqnarray}
for the gauge fields ($i = 1,2,3$), together with Gauss' law, which
here is a constraint derived from the gauge choice and used to
test the stability of the code,
\begin{equation}
2 \left( \psi_1 \stackrel{\leftrightarrow}{\partial_0} \psi_2 +
        \psi_3 \stackrel{\leftrightarrow}{\partial_0} \psi_4 \right)+
        \partial_i \dot{A}_i = 0 \,.
\end{equation}
The arrows indicate asymmetric derivatives.

This system is discretized using a standard staggered leapfrog method; however,
to reduce its relaxation time we also add an {\it ad hoc} dissipation term to
each equation ($\eta \dot{\psi}_i$ and $\eta \dot{A}_i$ respectively).  This is 
to allow the strings to `condense out' and be
identified.  In an expanding Universe the expansion rate would play such a role,
though $\eta$ would typically not be constant.  We tested a range of strengths
of dissipation, and checked that it did not significantly affect the number
densities obtained.  Further, our results are always compared to the cosmic
string case where dissipation would have the same effect.  The simulations we
display later used $\eta = 0.5$.

We set up initial conditions for the scalar field by placing the field 
in vacuum with random phases on a subgrid. We then iteratively interpolate by 
bisection onto the full grid; after each bisection the field is shifted into the 
vacuum before the next bisection takes place.

Having fixed the initial scalar field configuration, we need to
set the initial gauge field. One possibility is to choose the
minimum energy configuration on the (fixed) scalar field
background (this does not mean that the gauge fields will cancel all scalar 
gradients, as there are insufficient gauge degrees of freedom).
This can only be done exactly numerically, as in
Ref.~\cite{ABL97}, but in that paper it was also shown that an
approximate analytic minimization ignoring the magnetic flux term
performs perfectly well:
\begin{equation}
\label{analmin}
A_i({\bf x}) = \psi_1 \partial_i \psi_2 - \psi_2 \partial_i \psi_1 +
        \psi_3 \partial_i \psi_4 - \psi_4 \partial_i \psi_3 \,. 
\end{equation}

All field momenta are set to zero initially.

These initial conditions are one possible choice out of an infinity of possible 
ways in which one might try to represent conditions resembling a thermal 
phase transition. Rather than attempt a highly-accurate description of the 
transition, which is not necessary since we are aiming at order-of-magnitude 
estimates of the formation rate, we need only be confident that reasonable 
changes to the initial conditions will not significantly alter the results. This 
we have already tested in two-dimensional simulations \cite{ABL98}, where we 
considered a variety of initial conditions, including ones which may be closer 
to the sort of thermal environment considered in Ref.~\cite{YB}.

\section{The formation rate}

To minimize any systematic errors in our analysis, we always compare
our results to the case of cosmic strings, which is obtained by simply
ignoring one of the (complex) scalar fields, setting $\psi_3 = \psi_4
= 0$. This makes the defects topological, and the flux tubes formed
now map out the locations of winding in the scalar field
configurations. In these field theory simulations of cosmic strings,
we can follow the early stages of cosmic string network evolution by
displaying the density of magnetic flux, and we do the same with the
semilocal string simulations, as was done in two dimensions in
Ref.~\cite{ABL97}.

\begin{figure}[t!]
\centering 
\leavevmode\epsfysize=7.8cm \epsfbox{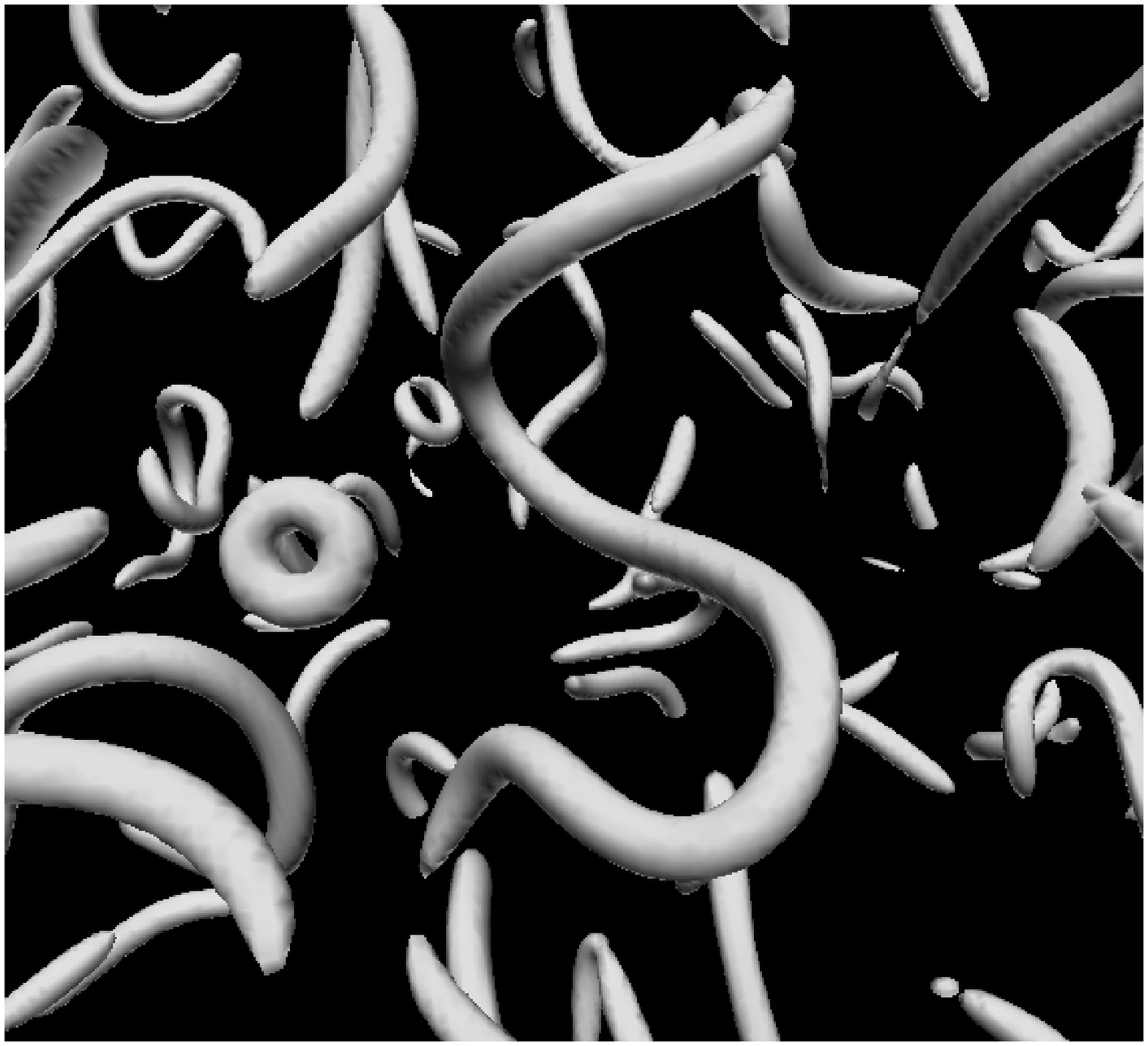}\\ % B&W
\vspace*{5pt}
\leavevmode\epsfysize=7.8cm \epsfbox{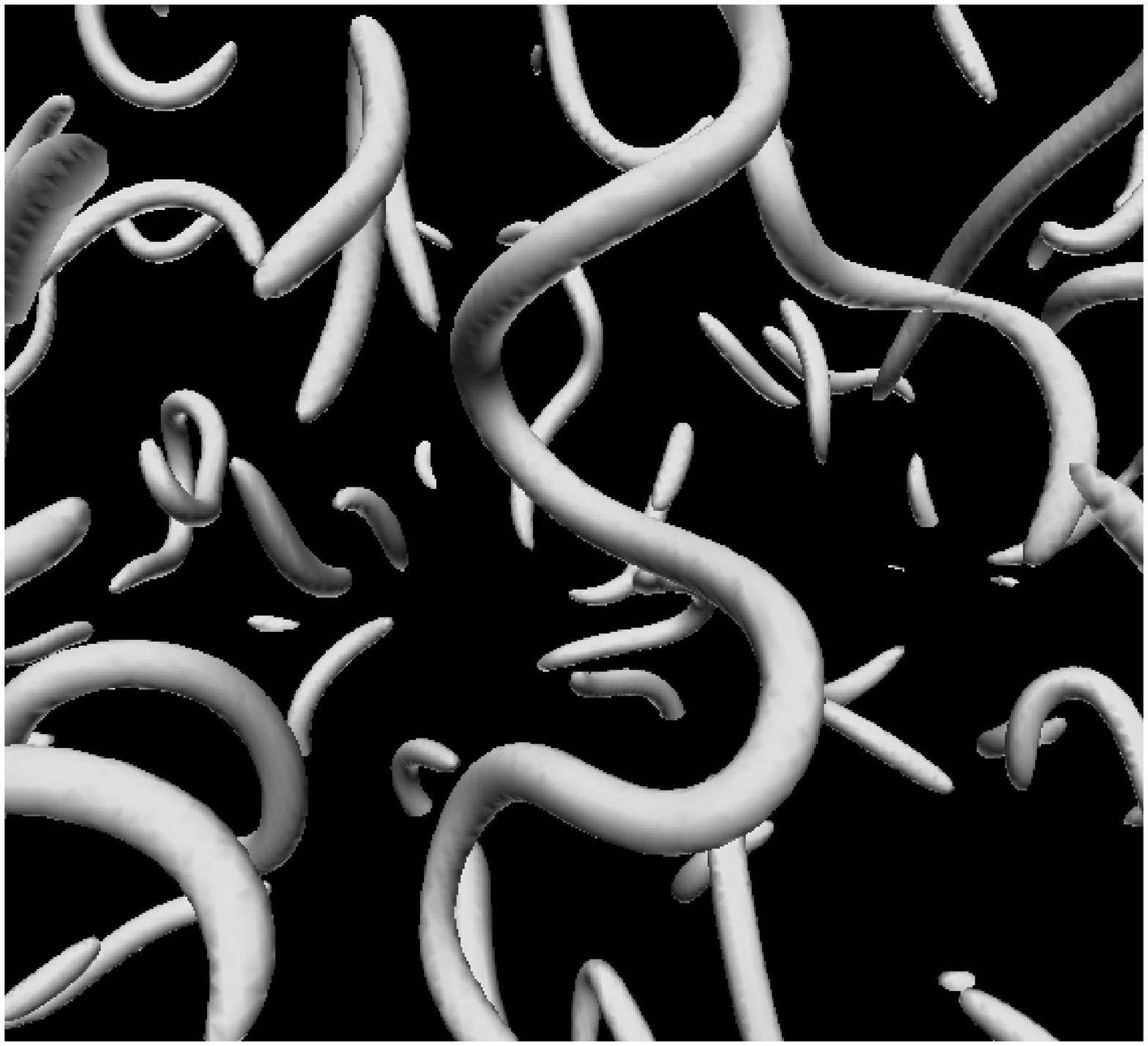}\\ % B&W
\vspace*{5pt}
\caption[semi1]{\label{semi1} Part of the large simulation, shown at time 60 
and time 70. Note several joinings of string segments, e.g.~two separate 
joinings on the long central string, and the disappearance of some loops. The 
different apparent thickness of strings is entirely an effect of 
perspective.}
\end{figure} 

The starting configurations obtained by our described procedure
initially yield a complicated mess of flux. However, after a few
timesteps this resolves itself into loops and open segments of
string. We observed a clear interaction between nearby segments which
join to form longer segments. Fig.~1 shows two time slices from a
single large $\beta = 0.05$ simulation, in a $256^3$ box, carried out on 
the Cray T3E at NERSC; these are
close-ups showing only part of the simulation box.\footnote{\footnotesize
Colour images and animations can be found on the WWW at {\tt
cfpa.berkeley.edu/$\sim$borrill/defects/semilocal.html}} We see
a collection of short string segments and loops; visually this is very
different from a cosmic string simulation where strings cannot have
ends. As time progresses, the short segments either disappear or link
up to form longer ones.

We can immediately conclude from these images that the formation rate
of semilocal strings is not extremely close to zero; the fact that
flux tubes are observed in our simulations implies that the formation
rate cannot be much smaller than one per correlation volume.

Nagasawa and Yokoyama \cite{NY} studied the related case of
electroweak defects and concluded that the initial density would be
totally negligible. However, this is not in contradiction with our
results because our semilocal strings arise during the evolution
due to back-reaction on the gauge fields from the scalar field
gradients. This enables initially short pieces of string to join up to
form semilocal strings of reasonable size, an effect not included in
their analysis.  Further, the electroweak string resides in a part of
parameter space where the defects are dynamically unstable, and in
this case we find that all the flux dissipates soon after the phase
transition.

\begin{figure}[t]
\centering 
\leavevmode\epsfysize=6.5cm \epsfbox{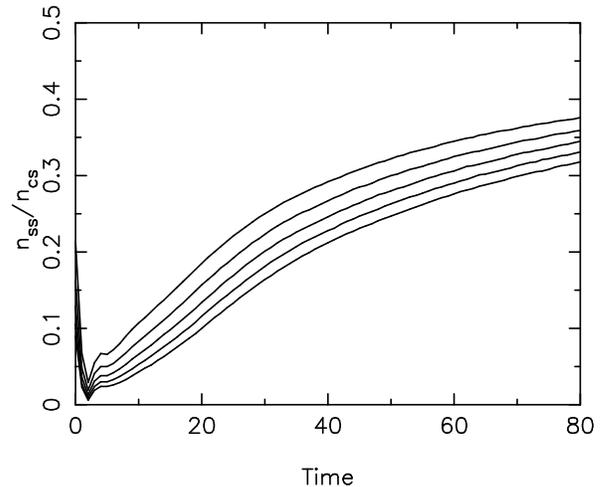}\\ 
\caption[semi2]{\label{semi2} This shows the ratio of total string
lengths in a semilocal and cosmic string simulation, with $\beta =
0.05$. The different lines show different magnetic flux thresholds,
from bottom to top they are 0.6, 0.55, 0.5, 0.45 and 0.4 times the
peak flux of a Nielsen--Olesen vortex.}
\end{figure} 

In order to quantify the formation rate, we compute the total length of string
in the simulations, always comparing the semilocal string density to that of a
cosmic string simulation with the same properties (including dissipation).  We
determine the length by setting a magnetic flux threshold and computing the
fractional volume of the box which exceeds it.  In Fig.~2, we plot the length of
semilocal string relative to the length found in cosmic string simulations, as a
function of time and with $\beta = 0.05$.  We see that after a transient during
which the initial tangle of flux sorts itself out, the system settles down to a
reasonable equilibrium. During that initial period the ratio of semilocal 
strings to cosmic ones grows, as cosmic ones are there right from the start due 
to topology while the semilocal ones need time to form. Even at late times there 
is a modest upward trend; we identify this as being caused by the
periodicity of the simulation box, which freezes-in any string crossing the box,
favouring cosmic string annihilation because of their higher density.  We take
the relative densities of semilocal and cosmic strings to be that at time 50 in
these simulations.  We shall investigate this trend more thoroughly in future
work using large simulations.  There is a modest dependence on the choice of
flux threshold, and we set it at one-half the flux density of a Nielsen--Olesen
vortex.

\begin{figure}[t]
\centering 
\leavevmode\epsfysize=6.3cm \epsfbox{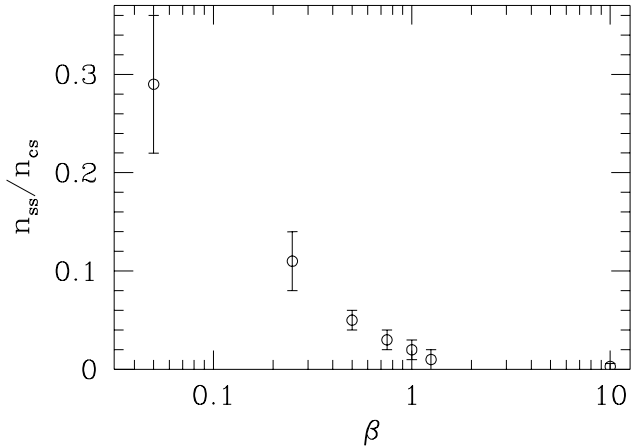}\\ 
\caption[semi3]{\label{semi3} The ratio of lengths of semilocal and cosmic 
strings.}
\end{figure} 

Fig.~3 shows the ratio of semilocal and cosmic string lengths, as a function of
the stability parameter $\beta$.  These results are derived from 700 simulations
(50 semilocal and 50 cosmic at each of 7 $\beta$ values) carried out in boxes of
dimension $64^3$ using a Sun Ultra II workstation.  Although the initial
correlation length of 16 units is a sizeable fraction of the box size, we are
only interested in brief evolution to allow the strings to become identifiable
and so boundary effects are not important.  The error bars include the
statistical spread between simulations, and an estimated 25\% systematic error
from the length counting algorithm (see the spread in Fig.~2) and the viscosity.
Those latter uncertainties are the dominant ones.  Recalling that the formation
rate of cosmic strings is estimated to be of order one per correlation volume
(0.88 in Ref.~\cite{VV}), these results are in excellent agreement with the
two-dimensional results we found in Ref.~\cite{ABL97}.  They show a significant
formation rate for low $\beta$, decreasing dramatically as $\beta \to 1$, beyond
which there is no evidence of semilocal string formation.  The small amount of
string seen in some large-$\beta$ simulations is an artifact of the viscosity;
the flux all dissipates if the viscosity is turned off, while it persists if
$\beta <1$.

\section{Conclusions}

Our simulations have shown that dynamically stable non-topological defects, such
as semilocal strings, can form with substantial density; this happens even if
the configuration immediately after the phase transition has no symmetry
restoration (recall our initial conditions place the scalar field in the vacuum
everywhere), and is due to the back-reaction of the gauge fields on the scalars.
For $\beta = 0.05$ we found that the formation rate was around 0.3 times that of
cosmic strings, while as $\beta$ is increased towards the stability/instability
transition at $\beta = 1$, the density drops to zero.

Moreover we have observed short segments of string joining to
form long strings and loops, in agreement with Refs.~\cite{H,GORS}, and
collapsing longitudinally, as invoked in some baryogenesis scenarios
\cite{BD}. It is important to study such phenomena in the related case of 
electroweak strings, where although the monopoles are of finite size their cores 
can become comparable to the separation as $\sin^2 \theta_{{\rm W}} \rightarrow 
1$. The details of these dynamics and their implications
deserve further study.

%%%%%%%%%%%%%%%%%%%%%%%%%%%%%%%%%%%%%%%%%%%%%%%%%%%%%%%%%%%%%%%%%%%%%%
\section*{Acknowledgments}

A.A.~acknowledges NSF grant PHY-9309364, CICYT grant AEN96-1668 and
UPV grant 063.310-EB225/95. J.B.~was supported by the Laboratory
Directed Research and Development Program of Lawrence Berkeley
National Laboratory under the U.S.~Department of Energy, Contract
No.~DE-ACO3-76SF00098, and used resources of the National Energy
Research Scientific Computing Center, which is supported by the Office
of Energy Research of the U.S.~Department of Energy. A.R.L.~was
supported by the Royal Society. We thank Kevin Campbell and the NERSC
Visualization Group, Graham Vincent at Sussex, and Konrad Kuijken and
the Kapteyn Institute for help with data visualization. A.R.L.~thanks
the University of Groningen for hospitality, and we thank Robert
Brandenberger, Mark Hindmarsh and Mark Trodden for useful discussions. 

%%%%%%%%%%%%%%%%%%%%%%%%%%%%%%%%%%%%%%%%%%%%%%%%%%%%%%%%%%%%%%%%%%%%%%

%%%%%%%%%%%%%%%%%%%%%%%%%%%%%%%%%%%%%%%%%%%%%%%%%%%%%%%%%%%%%%%%%%%%%%
\end{document}